# Theoretical study of N-complexes in carbon nanotubes

*Mariana Rossi, Adalberto Fazzio, Antônio J. R. da Silva*[†]

Instituto de Física da Universidade de São Paulo, São Paulo, SP, Brazil.

[†]ajrsilva@if.usp.br.

We have made a systematic theoretical study to determine which are the most stable structures for substitutional nitrogen defects in carbon nanotubes, by making total energy calculations via DFT. These calculations were made for a (5,5) and a (8,0) nanotube to check the influence of chirality and for a graphene sheet, to check the limit of infinite diameter. In all these systems, we found that either substitutional nitrogen or structures containing two vacancies surrounded by four pyridine-like rings have the lowest formation energy, depending on the value of the nitrogen chemical potential. These defects have lower formation energies in nanotubes than in the graphene sheet.

Carbon nanotubes have proved to be one of the structures with higher potential to be used in a wide variety of nano-mechanical or electronic applications[1-8]. Chemical doping of carbon nanotubes enhances their already outstanding properties. It allows, for example, the fabrication of diodes and transistors with this material, since these devices require *n*- and *p*-doping. By introducing new levels close to the Fermi level, chemical doping can also change the sensibility of carbon nanotubes to different kinds of molecules, making possible the fabrication of sensors [9,10].

Nitrogen doped CNT's ($CN_x$) have been reported to show greater sensibility to gases, such as ammonia or ethanol[9]. Many authors[14-23] made previous works on this subject and proposed geometries for the N doped nanotubes, but no systematic study was performed. Our proposal here is to systematicaly study the possible geometries for substitutional nitrogen in carbon nanotubes, in order to find the most stable and therefore, the most likely structures that may occur in a sample.

We studied a *zigzag* (8,0) nanotube and an *armchair* (5,5) nanotube to check the influence of chirality in the formation energies of the defects. Since these are relatively small diameter tubes, we also studied the graphene sheet, which stands for the infinite diameter limit. In all these systems, we found that either substitutional nitrogen or structures containing two vacancies surrounded by four pyridine-like rings have the lowest formation energy, depending on the value of the nitrogen chemical potential.

Our results are based on *ab initio* spin-polarized total energy density functional theory calculations [24,25] (SIESTA code[26]; GGA approximation[27]; pseudopotentials[28]; DZP basis set[29]; real space grid associated to an energy cutoff of 200 Ry; residual forces smaller than 0.03 eV/Å after structural optimizations). We used the supercell method with periodic boundary conditions (lateral sizes of 17.5 Å to prevent any direct interaction between the tubes images).

Different numbers of repetitions of the unitary cells were used: seven for the (5,5) (140 atoms) and five for the (8,0) nanotube (160 atoms); (9x9) for the graphene sheet (162 atoms). Regarding the nanotubes, three **k**-points in the direction of the axis of the tube and one in the other two directions were used for the Brillouin zone integration. For the graphene sheet, four **k**-points were used in the two directions parallel to the sheet and one in the perpendicular direction. All the atoms in the whole supercell were relaxed for the pristine systems in order to find the optimum lattice constants, which were 2.494 Å and 4.326 Å for the armchair and zigzag nanotubes, respectively, and 2.470 Å for the graphene sheet. However, when the defects were studied, we kept the atoms in one of the unitary cells of the tubes fixed. As we are interested in the limit of isolated defects, this was done in order to minimize the elastic interaction between them. This choice did not affect in a significant way the formation energies. When the structures were calculated without the constraints, the total energies did not change by more than 0.1 eV.

In order to study the $CN_x$ nanotubes, we calculated the formation energy $E_{form}$ of the defects, since this is a quantity that can be compared between systems with different numbers of atoms. These energies were calculated in the following way:

$$E_{form} = E_{sys}^{tot} + n_C \mu_C - E_{p-sys}^{tot} - n_N \mu_N \qquad (1)$$

In Equation 1, $E_{sys}^{tot}$ is the total energy of the nitrogen doped system, $E_{p-sys}^{tot}$ is the total energy of the pristine nanotube or the graphene sheet, depending on the system studied, $\mu_C$ and $\mu_N$ are the chemical potentials of the carbon and nitrogen atoms respectively, $n_C$ the number of carbon atoms removed from the system and $n_N$ the number of nitrogens inserted into the system.

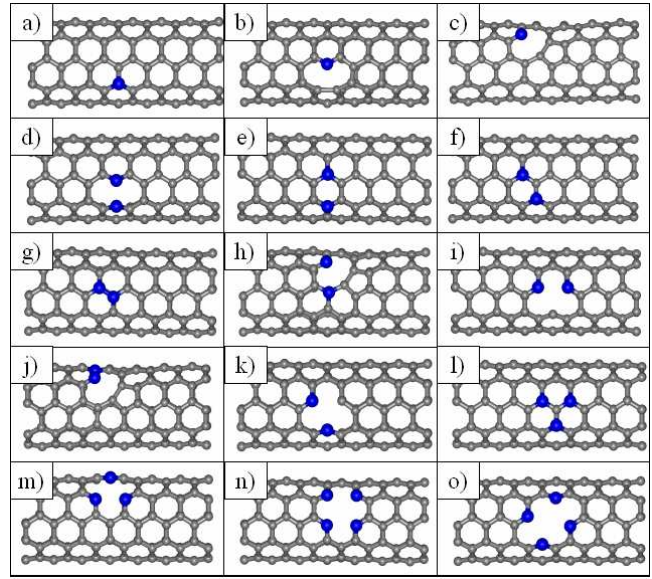

**Figure 1**. The final relaxed geometries for the 15 structures. The darker and bigger spheres correspond to nitrogens whereas the others correspond to carbons.

The chemical potential for carbon was taken from a graphene sheet, calculated as $E_{p-graf}^{tot}/N_C$, where $N_C$ is the number of carbons in the supercell (in our case, $N_C = 162$). If we had used the carbon nanotubes as the source of C atoms, the chemical potential would be higher by approximately 0.2eV, which would not change our conclusions. In the case of the nitrogen atoms, we considered many different possible reservoirs, starting with $N_2$ (which accounts for the highest binding energy for nitrogen, i.e. the lowest chemical potential). In order to check if the values considered for $\mu_N$ were realistic, we calculated the chemical potential of nitrogen in the ammonia molecule and the pyridine molecule. These values were found to be 1.2 eV and 0.9 eV above $N_2$'s chemical potential, respectively. The molecules used in the real experiments[11,12] are very complex and even though we have not calculated the N binding energy in them, they should have higher values than the ones we calculated.

First we studied 15 structures in the (5,5) tube (see Figure 1), and their formation energies are shown in Table 1.

**Table 1**. Formation energy for each of the defects presented in Figure 1 for $\mu_N = \mu_N(N_2)$. The structures in the first and third columns are labeled as in Fig. 1.

| Structure | $E_{form}$ (eV) | Structure | $E_{form}$ (eV) |
|---|---|---|---|
| a) | 0.51 | i) | 3.99 |
| b) | 3.96 | j) | 3.52 |
| c) | 3.37 | k) | 4.16 |
| d) | 1.25 | l) | 2.51 |
| e) | 1.19 | m) | 1.56 |
| f) | 1.42 | n) | 2.03 |
| g) | 2.14 | o) | 1.37 |
| h) | 3.38 | | |

For each family of defects we studied, shown in Figure 1, we plotted the formation energy of its most stable configuration against the nitrogen chemical potential (Figure 2).

In Figure 2 we can see that the defects with lower formation energy are structure **a** for lower nitrogen chemical potencial, and structures **m** and **o** for higher nitrogen chemical potentials. We notice, as well, that the structure proposed by the experimentalists (structure **m**) does not have the lowest formation energy for any value of $\mu_N$, implying that it will not be as abudant in a sample as the structure containing two vacancies surrounded by four pyridine-like rings (structure **o**).

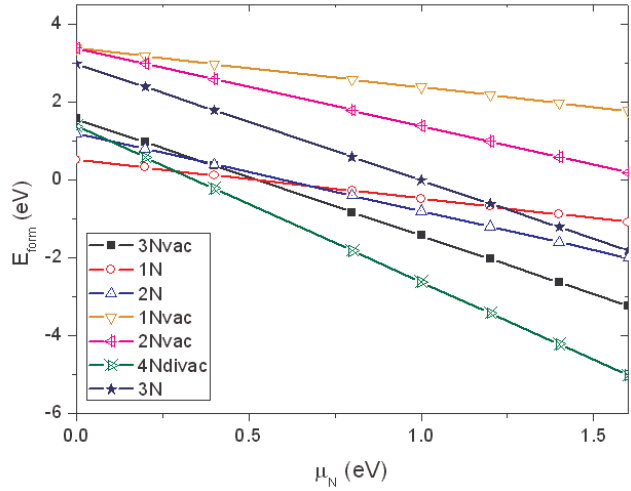

**Figure 2**. Formation energy of the defects in the (5,5) nanotube for varying nitrogen chemical potentials.

In order to check if this result was not valid only for (5,5) nanotubes, we calculated the formation energy for the most stable defects in a (8,0) nanotube and in the graphene sheet. These systems were chosen because the (8,0) nanotube is a *zigzag* tube, which should account for chirality effects and the graphene sheet accounts for the infinite diameter limit.

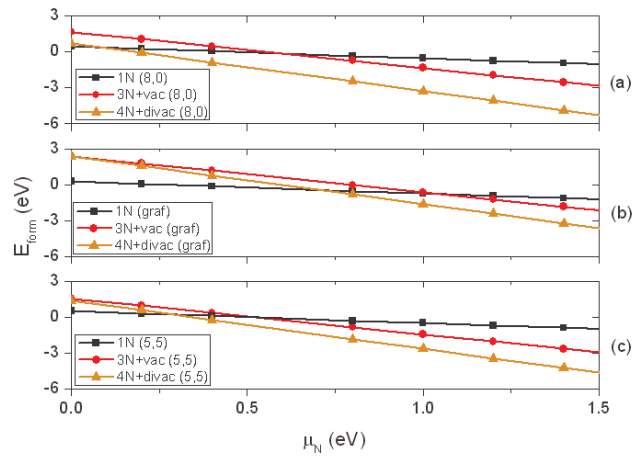

**Figure 3**. Formation energy of the most stable defects in: (a) the (8,0) nanotube; (b) the graphene sheet; (c) the (5,5) nanotube.

The lowest formation energy defects do not change in all systems studied, being either the substitutional N atom or the one with four nitrogens and two vacancies (4N-divac defect), as can be seen in Figure 3. In the nanotubes, the transition from substitutional nitrogen to the 4N-divac defect occurs for chemical potentials in the range of 0.1 to 0.25 eV above the value for $N_2$ as the source of N atoms. This is a condition easily met in the experiments. For graphene, even though the stability sequence does not change, this transition happens at higher $\mu_N$, at a value of approximately 0.6eV. This indicates that curvature effects tend to favor the removal of C atoms and the introduction of N atoms.

Note that the formation energies for the 4N-divac defect is negative for a large range of $\mu_N$, indicating that it is relatively easy to incorporate nitrogens in tubes, being almost impossible to prevent their incorporation.

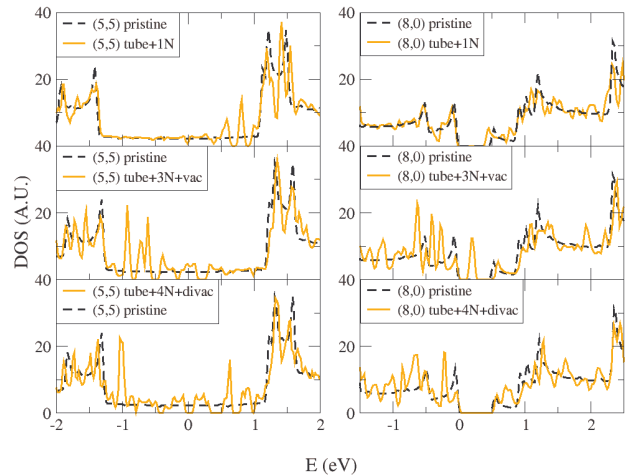

**Figure 4**. Density of states for the most stable defects (1N, 3N+vac and 4N+divac) in the nanotubes. For the (8,0) nanotube, the densities are aligned by the top of the valence band.

In Figure 4, we present the total density of states for the most stable defects in the (5,5) metallic nanotube and the (8,0) semiconductor nanotube. It is important to notice that the defect composed by three pyridine-like rings surrounding a vacancy introduces new mid gap states in the semiconductor tube, whereas the substitutional nitrogen (1N defect) introduces donor states close to the bottom of the conduction band, in accordance to previous works [16,21]. The 4N-divac defect, on the other hand, does not introduce any new gap states.

For a better characterization of the levels introduced by the nitrogens in the tubes, images of the charge density of particular DOS peaks are presented in Figures 5 and 6. In Figure 5 (a) the image corresponds to the unnocupied DOS peaks of the (5,5) 1 nitrogen defect located at 0.5 eV that can be seen in Figure 4; in Figure 5 (b) and (c) the results correspond to the occupied prominent peaks located at -1.0 eV that can also be seen in Figure 4 in the graphics that corresponds to the three and four nitrogen defects in the (5,5) tube, respectively.

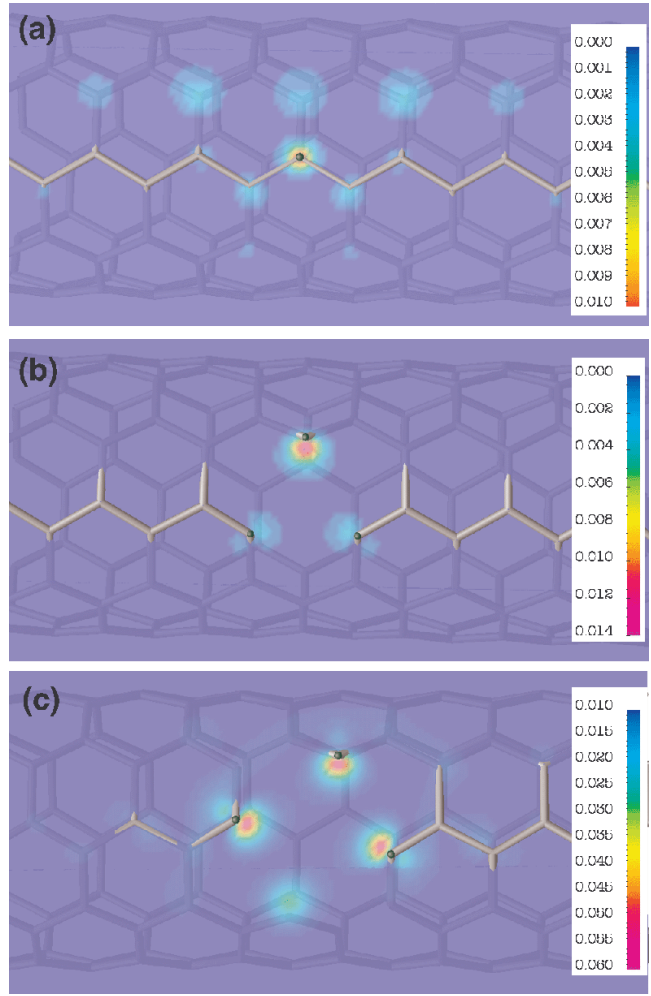

**Figure 5**. Charge density of particular DOS peaks (see text for details) for the (5,5) nanotube, in the region where the nitrogen states are localized. The planes of the images are tangent to the tube through one of the nitrogens (marked darker in the Figure). Figure (a) stands for the 1 nitrogen defect, (b) for the three nitrogen plus vacancy defect and (c) for the four nitrogen plus two vacancies defect. The scales are in $e^-/bohr^3$.

In Figure 6 we follow the same order of defects but now for the (8,0) tube. In Figure 6 (a), (b) and (c) the images correspond to the states located at 0.5 eV, 0.25 eV and -0.5 eV for the one, three and four nitrogens defect, respectively. They can also be observed in Figure 4.

We can see that the states related to the nitrogens are more localized in the (5,5) metallic nanotube than in the (8,0) semi-conductor tube.

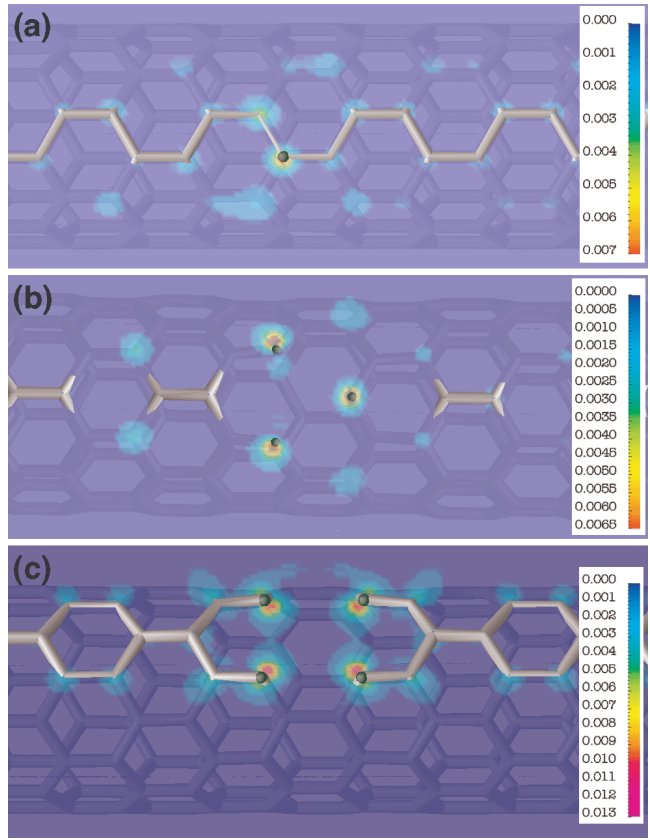

**Figure 6**. Charge density of particular DOS peaks (see text for details) for the (8,0) nanotube, in the region where the nitrogen states are localized. The planes of the images are tangent to the tube through one of the nitrogens (marked darker in the Figure). Figure (a) stands for the 1 nitrogen defect, (b) for the three nitrogen plus vacancy defect and (c) for the four nitrogen plus two vacancies defect. The scales are in e$^-$/bohr$^3$.

The 4N-divac structure is in agreement with the EELS and XPS spectra obtained by Terrones *et al.* [20,21], which show peaks related to pyridinic nitrogen. Another work with similar experimental results was made by Bulusheva *et al* [13]. The eventual differences between the theoretical and experimental peaks may be due to the concentration of the defects. It is necessary, however, to obtain more specific experimental results, such as vibrational spectra, in order to clearly identify the structure of these N defects.

Finally, we should comment that it is not possible to rule out the existence of more complex structures, perhaps containing even more nitrogens, but this study is beyond the scope of the present work.

The authors acknowledge Brazilian agencies FAPESP and CNPq for financial support and CENAPAD-SP for computational time.

**Table of Content Graphic**

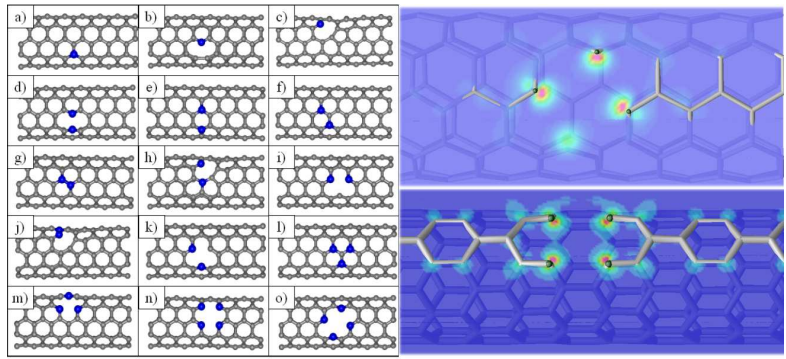